\documentclass[useAMS,usenatbib]{mn2e}
\usepackage{graphicx}
\usepackage{aas_macros}

\newcommand{\kms}{\mbox{${\rmn{km}\,\rmn{s}}^{-1}$}}

\newcommand{\Msolar}{\mbox{${M}_{\sun}$}}
\newcommand{\Lsolar}{\mbox{${L}_{\sun}$}}
\newcommand{\Rsolar}{\mbox{${R}_{\sun}$}}
\newcommand{\Vsini}{\mbox{${\rmn{V}}_{\rmn{rot}}\sin i$}}
\newcommand{\logg}{\mbox{$\log\,{\rmn{g}}$}}
\newcommand{\Teff}{\mbox{T$_{\rmn{eff}}$}}

\title[WASP 1628+10]{WASP 1628+10 -- an EL CVn-type binary with a very-low-mass
stripped-red-giant star and multi-periodic pulsations.} 
\author[P. F. L. Maxted et~al.]{P. F. L.~Maxted$^{1}$\thanks{E-mail:
p.maxted@keele.ac.uk}, A. M. Serenelli$^2$, T. R. Marsh$^3$,
S. Catal\'{a}n$^3$, D. P. Mahtani$^{1}$, \newauthor
V. S. Dhillon$^4$ \\
$^1$Astrophysics Group,  Keele University, Keele, Staffordshire ST5 5BG, UK\\
$^2$Instituto de Ciencias del Espacio (CSIC-IEEC), Facultad de Ciencias,
Campus UAB, 08193, Bellaterra, Spain.\\
$^3$Department of Physics, University of Warwick, Coventry CV4 7AL, UK\\
$^4$Department of Physics and Astronomy, University of Sheffield, Sheffield S3
7RH, UK
}

\date{To be inserted}
\pagerange{\pageref{firstpage}--\pageref{lastpage}} \pubyear{2014}

\begin{document}

\maketitle

\label{firstpage}

\begin{abstract}
 The star 1SWASP\,J162842.31+101416.7 (WASP\,1628+10) is one of several
EL~CVn-type  stars recently identified using the WASP database, i.e., an
eclipsing binary star in which an A-type dwarf star (WASP\,1628+10\,A)
eclipses the remnant of a disrupted red giant star (WASP\,1628+10\,B). We have
measured the masses, radii and luminosities of the stars in WASP\,1628+10
using  photometry obtained in three bands (u', g', r') with the Ultracam
instrument and medium-resolution spectroscopy. The properties of the remnant
are well-matched by models for stars in a rarely-observed state evolving to
higher effective temperatures at nearly constant luminosity prior to becoming
a very low-mass white dwarf composed almost entirely of helium, i.e., we
confirm that WASP\,1628+10\,B is a pre-He-WD. WASP\,1628+10\,A appears to be a
normal A2\,V star with a mass of $1.36\pm0.05$\Msolar. By fitting models to
the spectrum of this star around the H$\gamma$ line we find that it has an
effective temperature  ${\rm T}_{\rm eff,A} = 7500\pm200$\,K and a metallicity
$[{\rm Fe/H}]=-0.3 \pm 0.3$. The mass of  WASP\,1628+10\,B is only
$0.135\pm0.02$\Msolar. The effective temperature of this pre-He-WD is
approximately 9200\,K. The Ultracam photometry of WASP\,1628+10 shows
variability at several frequencies around 40 cycles per day, which is typical
for $\delta$~Sct-type pulsations often observed in early A-type stars like
WASP\,1628+10\,A. We also observe frequencies near 114 cycles/day and 129
cycles/day, much higher than the frequencies normally seen in $\delta$~Sct
stars. Additional photometry through the primary eclipse will be required to
confirm that these higher frequencies are due to pulsations in
WASP\,1628+10\,B. If confirmed, this would be only the second known example of
a pre-He-WD showing high-frequency pulsations. \end{abstract}


\section{Introduction}
 1SWASP~J162842.31+101416.7 (WASP\,1628+10 hereafter) is one of several
million bright stars ($8\la V \la 13$)  that have been observed by the Wide
Angle Search for Planets, (WASP, \citealt{2006PASP..118.1407P}).
\citet{2014MNRAS.437.1681M} showed that this star is an  eclipsing binary star
with an orbital period of 0.72\,days that contains an A-type dwarf star
(WASP\,1628+10\,A) and the precursor of a helium white dwarf (pre-He-WD) with
a mass $\la 0.3\Msolar$ (WASP\,1628+10\,B). 

 Low-mass white dwarf stars ($M\la 0.35\Msolar$) are the product of binary
star evolution \citep{1993PASP..105.1373I, 1995MNRAS.275..828M}. Various
evolution channels exist, but they are generally the result of mass transfer
from an evolved main sequence star or red giant star onto a companion star.
Towards the end of the mass transfer phase the donor star will have a
degenerate helium core. This ``stripped red giant star'' does not have
sufficient mass to ignite helium, and so the white dwarf that emerges has an
anomalously low mass and is composed almost entirely of helium. For this
reason, they are known as helium white dwarfs (He-WDs).  If the companion to
the red giant is a neutron star then the mass transfer is likely to be stable
so the binary can go on to become a low mass X-ray binary containing a
millisecond pulsar. Several millisecond radio pulsars are observed to have
low-mass white dwarf companions \citep{2008LRR....11....8L}. Many He-WDs have
been identified in the Sloan Digital Sky Survey \citep{2007ApJ...660.1451K},
some with masses as low as 0.16\Msolar\ \citep{2012ApJ...751..141K}, and from
proper motion surveys \citep{2009A&A...506L..25K}.  Helium white dwarfs can
also be produced by mass transfer from a red giant onto a main sequence star,
either rapidly through unstable common-envelope evolution or after a
longer-lived ``Algol'' phase of stable mass transfer
\citep{1969A&A.....1..167R, 1970A&A.....6..309G, 2004A&A...419.1057W,
1993PASP..105.1373I, 2003MNRAS.341..662C, 2001ApJ...552..664N}. He-WDs may
also be the result of collisions in dense stellar environments such as the
cores of globular clusters \citep{2008ApJ...683.1006K}, or by tidal stripping
of a red giant star by a super-massive black hole \citep{2013arXiv1307.6176B}.

 The evolution of He-WDs is expected to be very different from more massive
white dwarfs.  If the time scale for mass loss from the red giant is longer
than the thermal timescale, then when mass transfer ends there will still be a
thick layer of hydrogen surrounding the degenerate helium core. The mass of
the hydrogen layer depends on the total mass and composition of the star
\citep{2004ApJ...616.1124N}, but is typically 0.001\,--\,0.005\Msolar, much
greater than for typical white dwarfs (hydrogen layer mass $<10^{-4}$\Msolar).
The pre-He-WD then evolves at nearly constant luminosity towards higher
effective temperatures while the hydrogen layer mass is gradually reduced by
stable shell burning of hydrogen via the CNO cycle. This pre-He-WD phase can
last several million years for lower mass stars with thicker hydrogen
envelopes. CNO fusion becomes less efficient towards the end of this phase so
the star starts to fade and cool. 

 The smooth transition from a pre-He-WD to a He-WD can be interrupted by one
or more phases of unstable CNO burning (shell flashes) for pre-He-WDs with
masses $\approx 0.2$\,--\,0.3\Msolar\ \citep{1975MNRAS.171..555W,
1999A&A...350...89D}. These shell flashes substantially reduce the mass of
hydrogen  that remains on the surface. The mass range within which shell
flashes are predicted to occur depends on the assumed composition of the star
and other details of the models \citep{2001MNRAS.323..471A}.   The cooling
timescale for He-WDs that do not undergo shell flashes is much longer than for
those that do because their thick hydrogen envelopes can support residual p-p
chain fusion for several billion years.

 \citet{2013Natur.498..463M} presented strong observational support for the
assumption that He-WDs are born with thick hydrogen envelopes.  They found that
only models with thick hydrogen envelopes could simultaneously match their
precise mass and radius estimates for both stars in the EL~CVn-type binary
WASP\,0247$-$25, together with other observational constraints such as the
orbital period and the likely composition of the stars based on their
kinematics. In addition, they found that the pre-He-WD WASP\,0247$-$25\,B is a
new type of variable star in which a mixture of radial and non-radial
pulsations produce multiple frequencies in the lightcurve near 250 cycles/day.
This opens up the prospect of using asteroseismology to study the interior of
this star, e.g., to measure its internal rotation profile.

 In this paper we present new spectroscopy of WASP\,1628+10 that we use to
confirm that WASP\,1628+10\,B is a pre-He-WD and new  photometry that suggests
this pre-He-WD also shows pulsations at more than one frequency.

\section{Observations}
 
\subsection{Spectroscopy}
 We obtained 41 spectra of WASP\,1628+10 on the nights 2013 May 20\,--\,24
using the Intermediate Dispersion Spectrograph (IDS) on 2.5-m Isaac Newton
Telescope at the Observatorio del Roque de los Muchachos on La Palma, Spain.
We used the H2400B grating and a 1.2\,arcsecond slit with the EEV10 charge
coupled device (CCD) detector to obtain spectra with a dispersion of
0.23\AA/pixel at 4350\AA. The resolution of the spectra estimated by fitting
the several lines in a calibration arc spectrum is 0.45\AA.  The unvignetted
portion of the CCD covers the wavelength range 4115\,--\,4635\AA. Spectra were
extracted using the optimal extraction algorithm of
\citet{1986PASP...98..609H} in the {\sc pamela} application distributed by the
Starlink project\footnote{\it starlink.jach.hawaii.edu}. Observations
of each star were bracketed with arc spectra and the wavelength calibration
established from these arcs interpolated to the time of mid-exposure. The
exposure times used were 600\,s or 900\,s, resulting in spectra with a typical
signal-to-noise ratio of 25\,--\,45 per pixel. The spectra were flux
calibrated using a spectrum of the star BD+28~4211 \citep{1990AJ.....99.1621O}
obtained using a wide slit.

\subsection{Photometry}
 We obtained photometry of WASP\,1628+10 using the multi-channel photometer
Ultracam \citep{2007MNRAS.378..825D} mounted on the William Herschel 4.2-m
telescope at the Observatorio del Roque de los Muchachos. Images of
WASP\,1628+10 and the comparison star TYC~964-558-1 were obtained
simultaneously through u', g' and r' filters with an exposure time of 0.55
seconds for the g' and r' images and 1.65 seconds for the u' images. Ultracam
is a frame-transfer device and we only read-out the data in windows around the
target and comparision star so the dead-time between the exposures in only
25\,ms. The image scale is 0.3 arcseconds per pixel. A log of the observations
is provided in Table~\ref{ObsTable}.

\begin{table}
 \caption{Observing log for our Ultracam observations of WASP\,1628+10.
``Night'' is the UTC date at the start of the night.
\label{ObsTable}}
 \begin{tabular}{lrrl}
\hline
  \multicolumn{1}{l}{Night} &
  \multicolumn{1}{l}{Start} &
  \multicolumn{1}{l}{End} &
  \multicolumn{1}{l}{Notes} \\
  \multicolumn{1}{l}{(UTC)} &
  \multicolumn{1}{l}{(UTC)} &
  \multicolumn{1}{l}{(UTC)} &
  \multicolumn{1}{l}{} \\
\hline
\hline
 \noalign{\smallskip}
 2013 04 21& 03:13 & 06:08 & Secondary eclipse \\
 2013 04 22& 00:09 & 06:01 & \\
 2013 04 23& 23:09 & 00:57 & Primary eclipse   \\
 2013 04 23& 03:04 & 05:59 & \\
 2013 04 24& 23:12 & 01:24 & \\
 \noalign{\smallskip}
\hline
 \end{tabular}   
 \end{table}     

 All reductions were performed with the Ultracam pipeline software. The images
were bias-subtracted and flat-fielded using twilight sky exposures in the
normal way. We used synthetic aperture photometry  to measure the apparent
flux of WASP\,1628+10 and the comparison star. The aperture radius was set to
twice the full-width at half-maximum (FWHM) of the stellar profile in each
image. The FWHM of the stellar profile was typically 1.5\,--\,2 arcsec.

\section{Analysis}
\subsection{Photometry}

\subsubsection{Pulsations}

\begin{figure}
\mbox{\includegraphics[width=0.49\textwidth]{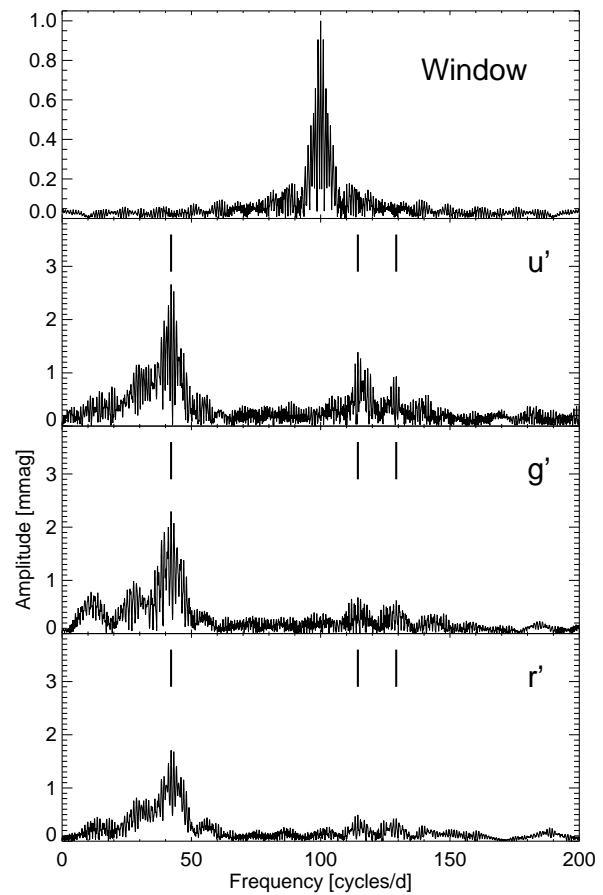}}
\caption{Periodograms of the out-of-eclipse Ultracam data of WASP\,1628+10. 
The three frequencies detected in all three channels are indicated. 
\label{fftfig}}
\end{figure}

 We used the {\sc period04} software package \citep{2005CoAst.146...53L} to
search for periodic variations in the three Ultracam datasets listed in
Table~\ref{ObsTable} that do not include an eclipse. For each dataset we first
removed data affected by clouds and then divided the differential magnitudes
by a low-order polynomial fit by least-squares. Periodograms were generated
for  the u', g' and r' data independently over the frequency range 0\,--\,1000
cycles/day in 0.017 cycle/day steps using the flux values in 10\,s bins
weighted by the standard error of the mean in each bin. The resulting
periodograms are shown in Fig.~\ref{fftfig}. We identified the frequency with
the highest peak in this periodogram and then performed a least-squares
fit to the data to optimise the values of the amplitude, phase and frequency
of this signal. We then identified the strongest frequency in the periodogram
using the residuals from this fit, and performed another least squares fit to
optimise all the frequencies, phases and amplitudes. This process was repeated
until we judged that the strongest frequency detected in the residuals was
due to noise.  The frequencies detected in each dataset and their amplitudes
are listed in Table~\ref{FreqTable}. 

 The comparision star we used has optical-infrared magnitudes entirely
consistent with those of a typical mid-K-type star. The other stars in the
Ultracam field of view are too faint to be useful as comparison stars. The
WASP lightcurve of the comparision star star has over 30,000 observations over
5 years and is constant to with 0.02 magnitudes. There is no reason to expect
that such a star would show variability  at the amplitudes and frequencies
seen in our Ultracam data.

 Some of the power at low frequencies ($\la 20$ cycles/day) in these data may
be due to offsets between data from different nights. The poor sampling of the
data also means that there are problems with 1-day aliases of real periodic
signals in the periodograms. Nevertheless, there are three frequencies seen in
the u', g' and r' data sets that agree to better than 0.5 per~cent. These are
indicated in Fig.~\ref{fftfig} and the mean value of these frequencies with
their standard errors are given in Table~\ref{FreqTable}. There are clearly
other frequencies in the range 25\,--\,50 cycles/day present in the data, but
we are unable to reliably identify the correct alias from the data currently
available.

\begin{table}
 \caption{Frequencies detected in our Ultracam data of WASP\,1628+10.
Amplitudes are given as the semi-amplitude of the fractional flux variation.
The standard error in the final digit of the mean value is given in
parentheses. The values listed here are the result of a least-squares fit to
the data to optimise the frequency, phase and amplitudes of all the
frequencies listed.
\label{FreqTable}}
 \begin{tabular}{rrrrcrrr}
\hline
  \multicolumn{4}{c}{Frequency (cycles/day)} &
  \multicolumn{4}{c}{Amplitude ($/10^{-3}$)} \\
  \multicolumn{1}{c}{u'} &
  \multicolumn{1}{c}{g'} &
  \multicolumn{1}{c}{r'} &
  \multicolumn{1}{c}{Mean}&
&
  \multicolumn{1}{c}{u'} &
  \multicolumn{1}{c}{g'} &
  \multicolumn{1}{c}{r'}\\
\hline
\hline
 \noalign{\smallskip}
 15.38 & 13.71 &        &           && 0.6 & 0.8 &      \\
 32.49 & 27.40 & 25.13  &           && 1.0 & 0.6 &  0.5 \\
 37.86 & 28.82 & 31.51  &           && 0.9 & 1.1 &  0.8 \\
       & 36.85 & 41.84  &           &&     & 1.0 &  0.7 \\
 42.13 & 42.11 & 42.20  & 42.15(3)  && 2.4 & 2.3 &  1.7 \\
 44.36 & 43.53 &        &           && 1.0 & 1.2 &      \\
 114.4 & 114.4 &  114.4 & 114.40(1) && 1.3 & 0.7 &  0.4 \\
 129.2 & 129.2 &  129.2 & 129.23(1) && 0.9 & 0.5 &  0.4 \\
 \noalign{\smallskip}
\hline
 \end{tabular}   
 \end{table}     

\subsubsection{Eclipses\label{eclipses}}
\begin{figure}
\mbox{\includegraphics[width=0.47\textwidth]{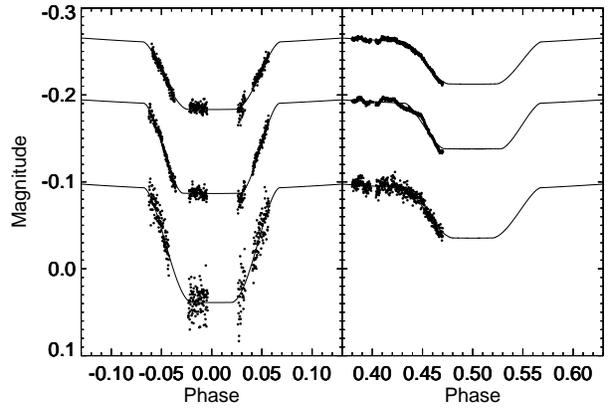}}
\caption{From bottom-to-top: Ultracam u', g' and r' lightcurves of the primary
and secondary eclipses of WASP\,1628+10 (points) with our  least-squares fits
to these data (lines). Note that the model fits shown do not include the
modulation of the signal due to pulsations.
\label{lcfitfig}}
\end{figure}

 Our Ultracam observations include the ingress to one secondary eclipse and
observations of one primary eclipse with gaps due to thin cloud. We used {\sc
jktebop}\footnote{\it www.astro.keele.ac.uk/$\sim$\it jkt/codes/jktebop.html}
version 28 (\citealt{2010MNRAS.408.1689S} and references therein) to analyse
the data from the u', g' and r' bands independently using the {\sc ebop}
lightcurve model \citep{1981psbs.conf..111E,1981AJ.....86..102P}. We used the
feature available in  {\sc jktebop} to modulate the flux from either star
using up to 5 sinusoidal functions, $A_i\sin[2\pi(t-T_i)f_i]$. We included in
our lightcurve model sinusoids at the 3 frequencies listed in Table~\ref{FreqTable}
detected in all three channels plus the 2 other sinusoids at the frequencies
with the largest amplitudes in each channel. The phase and amplitude of each
frequency were included as free parameters in the fit. We assumed that
frequencies $>100$ cycles/day originate from WASP\,1628+10\,B  and that lower
frequencies originate from WASP\,1628+10\,A. There is little difference to the
quality of the fit or the parameters derived if all the frequencies are
assigned to WASP\,1628+10\,A.  We used magnitude values in 10\,s bins with
equal weight for each binned data point for the least-squares fit. Other free
parameters in the least-squares fit were: a normalisation constant, the
surface brightness ratio $J = S_B/S_A$, where $S_A$ is the surface brightness
of WASP\,1628+10\,A\footnote{More precisely, {\sc jktebop} uses the surface
brightness ratio for the stars calculated at the centre of the stellar discs,
but for convenience we quote the mean surface brightness ratio here.} and
similarly for $S_B$; the sum of the radii relative to semi-major axis,
$s=(R_{\rm A}+R_{\rm B})/a$; the ratio of the radii, $k=R_{\rm B}/R_{\rm A}$;
the orbital inclination, $i$; the phase of primary eclipse, $\Delta$. The
orbital phase of the observations was calculated using the ephemeris for the
time of primary eclipse (min~I) derived from a similar lightcurve fit to the
WASP photometry of WASP\,1628+10 by \citet{2014MNRAS.437.1681M}, which we
quote here for convenience:
\[ {\rm HJD(UTC)~min~I} = 2454921.8532(5) + 0.7203633(9)\times E. \]
 The standard errors on the final digits quoted in parantheses imply a
standard error in the phase of the Ultracam observations of 0.0025. The mass
ratio was fixed at the value derived from the spectroscopy described below.
Linear limb-darkening coefficients for star A were taken from
\citet{2004A&A...428.1001C}. We assumed that the orbit is circular and fixed
the gravity darkening coefficients of both stars to 0.5 since this parameter
has a negligible effect on the shape of the eclipses. The linear limb
darkening coefficient of star B also has a negligible effect on the lightcurve
so was fixed at a value of 0.5. The results are given in
Table~\ref{lcfitTable}. The observed lightcurves and the model fits are shown
in Fig.~\ref{lcfitfig}. 

It is difficult to assign reliable error estimates to the parameters derived
from the individual lightcurves because of the large number of possible
pulsation frequencies that may be present in the data and the unknown effect
of the pulsations on the shape of the ingress and egress to the eclipses
\citep{2011MNRAS.416.1601B}. Instead, we simply quote the values derived from
each lightcurve plus the values derived by \citet{2014MNRAS.437.1681M} for the
same parameters using the WASP photometry and an r' lightcurve obtained with
the PIRATE telescope. We then adopt the mean values of the parameters that do
not depend on wavelength from these 5 lightcurve fits and use the standard
error on the mean to estimate the standard errors for these estimates. These
mean values and their standard errors are also given in
Table~\ref{lcfitTable}.

\begin{table*}
\caption{Parameters for the lightcurve models fit by least-squares. Parameters
fixed in the least-squares fit are preceeded by ``=''.  The luminosity ratio
in each band is  $\ell = L_B/L_A$, other  parameter definitions are given in
the text. $N$ is the number of points used in the least-squares fit and RMS is
the standard deviation of the residuals. Results for the fits to the PIRATE r'
and WASP lightcurves are from \citet{2014MNRAS.437.1681M}. HMJD is
heliocentric modified Julian date. }
\label{lcfitTable}
 \begin{tabular}{@{}llcrrrrrr}
\hline
&&& \multicolumn{3}{c}{Ultracam}&\multicolumn{1}{c}{PIRATE} 
& \multicolumn{1}{c}{WASP} \\
\multicolumn{2}{@{}l}{Parameter}
 & Units &
\multicolumn{1}{c}{u'}& 
\multicolumn{1}{c}{g'}& 
\multicolumn{1}{c}{r'}& 
\multicolumn{1}{c}{r'}& 
\multicolumn{1}{c}{--}&
\multicolumn{1}{c}{Mean} \\
\hline
\hline
$J      $&Surface brightness ratio           &&$ 2.23   $&$ 1.74   $&$ 1.79   $&$ 1.90  $&$  1.79 $&$J_{\rm r'}=1.85\pm0.06$\\
$s      $&Sum of fractional radii            &          &$ 0.495  $&$ 0.508  $&$ 0.499  $&$ 0.491 $&$  0.488$&$ 0.496\pm 0.004 $\\
$k      $&Ratio of the radii                 &          &$ 0.225  $&$ 0.230  $&$ 0.215  $&$ 0.209 $&$  0.227$&$ 0.221\pm 0.004 $\\
$u_A    $&Linear limb darkening coefficient  &          &$ =0.72  $&$ =0.68  $&$ =0.54  $&$ 0.31  $&$  0.84 $&$                $\\
$i      $&Orbital inclination                 &$^{\circ}$&$ 73.4   $&$ 73.3   $&$ 72.4   $&$ 72.9  $&$  73.7 $&$ 73.1 \pm 0.2   $\\
$\Delta $&Phase offset of primary eclipse    &          &$ 0.0036 $&$ 0.0053 $&$ 0.0024 $&$       $&$ $&$0.0038\pm0.0008 $\\
$\ell   $&Luminosity ratio                   &&$ 0.137  $&$ 0.109  $&$ 0.087  $&$ 0.083 $&$  0.094$&$\ell_{\rm r'}=0.085\pm0.002$\\
$R_A/a  $&Fractional radius of star A        &          &$ 0.404  $&$ 0.412  $&$ 0.411  $&$ 0.406 $&$  0.398$&$ 0.406\pm 0.003 $\\
$R_B/a  $&Fractional radius of star B        &          &$ 0.091  $&$ 0.095  $&$ 0.089  $&$ 0.085 $&$  0.090$&$ 0.090\pm 0.002 $\\
$f_1    $&                                   & d$^{-1}$ &$ =42.15 $&$ =42.15 $&$ =42.15 $&$       $&$       $&$                $\\
$T_1    $&                                   & HMJD     &$ 0.0036 $&$ 0.0009 $&$ 0.0009 $&$       $&$       $&$                $\\
$A_1    $&                                   & mmag     &$ -2.1   $&$ -2.5   $&$ -1.2   $&$       $&$       $&$                $\\
$f_2    $&                                   & d$^{-1}$ &$ =114.40$&$ =114.40$&$ =114.40$&$       $&$       $&$                $\\
$T_2    $&                                   & HMJD     &$ -0.0004$&$ -0.0012$&$ -0.0010$&$       $&$       $&$                $\\
$A_2    $&                                   & mmag     &$ 12.8   $&$ 12.4   $&$ 8.6    $&$       $&$       $&$                $\\
$f_3    $&                                   & d$^{-1}$ &$ =129.23$&$ =123.23$&$ =123.23$&$       $&$       $&$                $\\
$T_3    $&                                   & HMJD     &$ -0.0046$&$ -0.0004$&$ -0.0003$&$       $&$       $&$                $\\
$A_3    $&                                   & mmag     &$ 5.3    $&$ -6.0   $&$ -4.1   $&$       $&$       $&$                $\\
$f_4    $&                                   & d$^{-1}$ &$ =32.49 $&$ =43.53 $&$ =31.51 $&$       $&$       $&$                $\\
$T_4    $&                                   & HMJD     &$ -0.0228$&$ -0.0001$&$ 0.2287 $&$       $&$       $&$                $\\
$A_4    $&                                   & mmag     &$ -1.4   $&$ 2.1    $&$ 0.9    $&$       $&$       $&$                $\\
$f_5    $&                                   & d$^{-1}$ &$ =37.86 $&$ =28.82 $&$ =41.84 $&$       $&$       $&$                $\\
$T_5    $&                                   & HMJD     &$ -0.0016$&$ 0.0417 $&$ -0.0189$&$       $&$       $&$                $\\
$A_5    $&                                   & mmag     &$ -2.8   $&$ 1.2    $&$ -1.2   $&$       $&$       $&$                $\\
N        &                                   &          &$  3052  $&$  3094  $&$  3063  $&$  480  $&$30131  $&$                $\\
RMS      &                                   & mmag     &$  9.2   $&$  3.8   $&$  3.4   $&$  6    $&$   42  $&$                $\\
\hline
\end{tabular}
\end{table*}

\subsection{Spectroscopy}

\subsubsection{Radial velocity measurements}
 We first attempted to measure the radial velocities of WASP\,1628+10\,A using
cross-correlation against the spectrum of $\theta$~Leo, an  A2\,V-type star
with narrow lines, obtained with the same instrumental setup. We excluded the
broad H$\gamma$ line from the calculation of the cross-correlation functions
(CCFs) and measured the radial velocity from a parabolic fit to the three
highest points in the CCF. The resulting radial velocities show large scatter
and obvious offsets between spectra obtained at the same orbital phase on
different nights. We assume that these systematic errors are the result of
$\delta$ Scuti-type pulsations in WASP\,1628+10\,A. These pulsations will
cause distortions of the spectral lines \citep[e.g.,][]{1998ApJ...495..440K}.
Some evidence for this is provided by the observation that the measured radial
velocities show better agreement from night-to-night if we apply a Gaussian
smoothing algorithm to the template spectrum. This is presumably a result of
the broader line profiles in a broadened template being less sensitive to the
detailed shape of distorted line profiles.  To derive the semi-amplitude,
$K_{\rm A}$, from these radial velocities we use a least-squares fit of a
circular orbit, i.e., the function $V_r = \gamma_{\rm A} + K_{\rm
A}\sin[2\pi(t-T_0)/P]$ with $\gamma_{\rm A}$ and $K_{\rm A}$ as free
parameters and $T_0$ and $P$ fixed at the values given in
Section~\ref{eclipses}. This gives a good fit to the radial velocities
measured with a smoothed template spectrum, but the derived value of the
velocity semi-amplitude then shows a dependence on the width of the Gaussian
smoothing kernel used. 

\begin{figure}
\mbox{\includegraphics[width=0.49\textwidth]{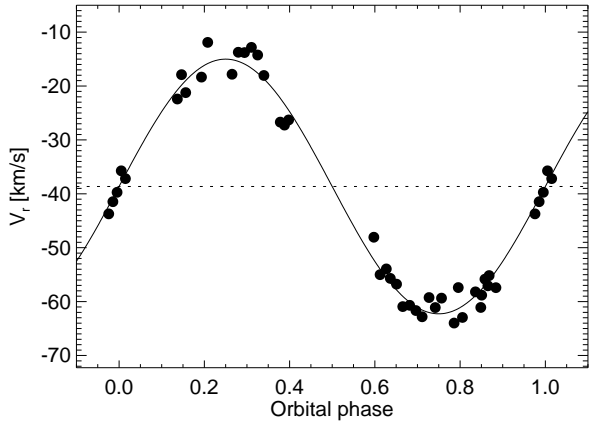}}
\caption{Radial velocities of WASP\,1628+10\,A, measured using
cross-correlation against a synthetic template spectrum generated by SME for
the parameters \Teff=7500\,K, \logg=4.2, \Vsini=105\kms. The solid line is a
circular orbit with the parameters $K_{\rm A}=23$\kms\ and the dotted line
indicates the mean radial velocity $\gamma = -39$\kms.
\label{RVFig}} \end{figure}

  We also measured the radial velocity of WASP\,1628+10\,A using a best-fit
synthetic spectrum derived from the analysis described in
section~\ref{smesect}. We found that the radial velocities derived varied
systematically depending on the method used to measure the position of the
CCF, e.g., a parabolic fit to the highest three points leads to a value of
$K_A$ lower by about 2.5\kms\ than the value derived from a Gaussian profile
fit to the region $\pm 100$\,\kms\ around the peak of the CCF.   The radial
velocities measured using this latter method are given in Table~\ref{RVTable}
and the least-squares fit of a circular orbit to these velocities is shown in
Fig.~\ref{RVFig}. 

 For this analysis we adopt the values $\gamma_{\rm A} = -39\pm3$\,\kms\ and
$K_{\rm A}=24\pm3$\,\kms, where the adopted values are from the least-squares
fit the the radial velocities given in Table~\ref{RVTable} and the estimated
standard errors reflect the range of different values derived from different
methods for measuring the radial velocity. 

\begin{table}
\caption{Measured radial velocities of  WASP\,1628+10\,A. These measurements
are affected by systematic errors due to pulsations and so we do not quote the
standard errors derived from the fitting procedure here.
\label{RVTable}}
\begin{tabular}{rrrr}
  \multicolumn{1}{l}{HJD(UTC)} & 
  \multicolumn{1}{c}{$V_r$} &
  \multicolumn{1}{l}{HJD(UTC)} & 
  \multicolumn{1}{c}{$V_r$} \\
  \multicolumn{1}{r}{$-2\,450\,000$} & 
  \multicolumn{1}{r}{[\kms]} &
  \multicolumn{1}{r}{$-2\,450\,000$} & 
  \multicolumn{1}{r}{[\kms]} \\
\hline
 \noalign{\smallskip}
6433.4478 &$  13.4$& 6435.4758 &$  21.7$ \\  
6433.4549 &$  12.8$& 6435.4863 &$  28.2$ \\  
6433.4620 &$  13.8$& 6435.5274 &$  22.3$ \\  
6433.6060 &$  -8.0$& 6435.5380 &$  26.3$ \\  
6433.6165 &$ -15.0$& 6435.5485 &$  26.3$ \\  
6433.6270 &$ -13.9$& 6435.5602 &$  27.2$ \\  
6434.4620 &$ -23.9$& 6435.5707 &$  25.8$ \\  
6434.4690 &$ -17.3$& 6435.5813 &$  22.0$ \\  
6434.4761 &$ -22.9$& 6436.5153 &$ -15.6$ \\  
6434.5070 &$ -21.0$& 6436.5258 &$ -16.7$ \\  
6434.5141 &$ -15.8$& 6436.5363 &$ -20.8$ \\  
6434.5212 &$ -15.1$& 6436.5480 &$ -20.6$ \\  
6434.5326 &$ -17.3$& 6436.5585 &$ -21.6$ \\  
6434.5986 &$  -3.6$& 6436.5690 &$ -22.7$ \\  
6434.6057 &$  -1.4$& 6436.5807 &$ -19.2$ \\  
6434.6128 &$   0.4$& 6436.5912 &$ -21.0$ \\  
6434.6198 &$   4.3$& 6436.6018 &$ -19.3$ \\  
6434.6269 &$   2.9$& 6436.6591 &$ -18.1$ \\  
6434.7146 &$  17.7$& 6436.6696 &$ -18.8$ \\  
6434.7217 &$  22.2$& 6436.6802 &$ -17.0$ \\  
6434.7287 &$  18.9$&                     \\
 \noalign{\smallskip}
\hline
\end{tabular}   
\end{table}     

\begin{figure}
\mbox{\includegraphics[width=0.47\textwidth]{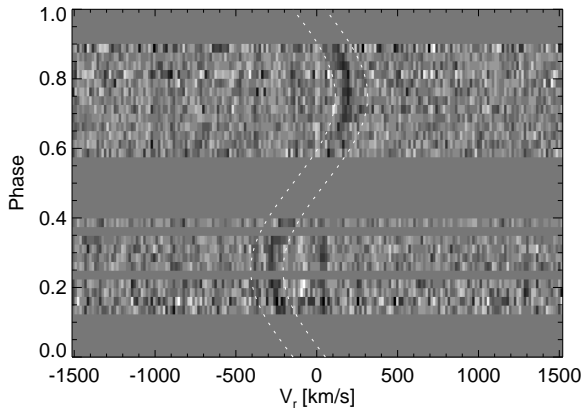}}
\caption{Grey-scale plot showing spectra of WASP\,1628+10 after subtraction of
a synthetic spectrum for WASP\,1628+10\,A. The sinusoidal signal highlighted
with dotted lines is the Mg\,II 4481 line from WASP\,1628+10\,B. The
absorption and emission features near 0\,\kms\ are residuals from the spectral
subtraction due to line profile variations in WASP\,1628+10\,A.
\label{trailMgIIfig}}
\end{figure}

 To measure the radial velocity of WASP\,1628+10\,B, we first subtracted a
scaled and shifted version of  a best-fit synthetic spectrum of
WASP\,1628+10\,A from all the spectra observed between orbital phases 0.1 and
0.9. This removes almost all of the signal of WASP\,1628+10\,A from these
spectra, revealing the spectrum of WASP\,1628+10\,B (Fig.~\ref{trailMgIIfig}).
Some weak features from the spectrum of WASP\,1628+10\,A remain in these
spectra because of the line profile variations caused by the pulsations in
this star. The Mg\,II~4481\AA\ line from WASP\,1628+10\,B can be clearly seen
in these spectra, but is difficult to measure in individual spectra. Instead,
we simultaneously fit a Gaussian profile to all these spectra with the
position of the line in the spectrum observed at time $t$ set by a radial
velocity offset $\gamma_{\rm B}+K_{\rm B} \sin(2\pi(t-T_0)/P)$. The width and
depth of the line are also free parameters in the least-squares fit. With this
method we obtain the values  $\gamma_{\rm B} = -41\pm 2$\,\kms\ and $K_{\rm B}
= -236\pm 3$\,\kms. The value of $\gamma_{\rm B}$ agrees well with the value
of $\gamma_{\rm A}$ derived from the radial velocities of WASP\,1628+10\,A.

\subsubsection{Disentangled spectra}
 Disentangling refers to the recovery of the individual spectra of two stars
from a series of combined spectra in which the individual spectra have
different radial velocity shifts due to the orbital motion in the binary
system. Here we use our own implementation of the algorithm by
\citet{1994A&A...281..286S} that uses a sparse matrix inversion technique to
recover the individual spectra. In outline, the problem we wish to solve is to
find the matrix {\boldmath {$x$}} that solves the equation
\mbox{\boldmath{$A\cdot x=b$}} in the least-squares sense,  where {\boldmath
{$b$}} is a column vector containing the observed spectra, {\boldmath{$x$}}
are the two individual spectra and {\boldmath{$A$}} performs a linear
interpolation of {\boldmath{$x$}} onto {\boldmath{$b$}} accounting for the
known radial velocity shifts of the two stars.  We have modified this
algorithm to account for the fact that some of our spectra were observed
during the total eclipse of WASP\,1628+10\,B. This is advantageous because it
removes the ambiguity over the contribution of each star to the combined
continuum level. The modification is straightforward since it only requires
the matrix elements of {\boldmath{$A$}} to by multiplied by the fractional
contribution of each star to the combined spectrum, i.e., $0$ in the case of
the matrix elements corresponding to WASP\,1628+10\,B observed during the
primary eclipse. To find the optimum luminosity ratio of the two stars in the
spectra observed outside of eclipse, $\ell$, we calculated the standard
deviation of the residuals ($\sigma$) over a grid of  $\ell$ values, and then
interpolated to the value of $\ell = 0.132$ that minimizes $\sigma$.  This
value is in good agreement with the value of $\ell$ we would expect based on
the fits to the u'- and g'-band lightcurves (Table~\ref{lcfitTable}).

\subsection{Effective temperature estimates}

\subsubsection{SME fits to the disentangled spectra\label{smesect}}
\begin{figure*} \mbox{\includegraphics[width=0.97\textwidth]{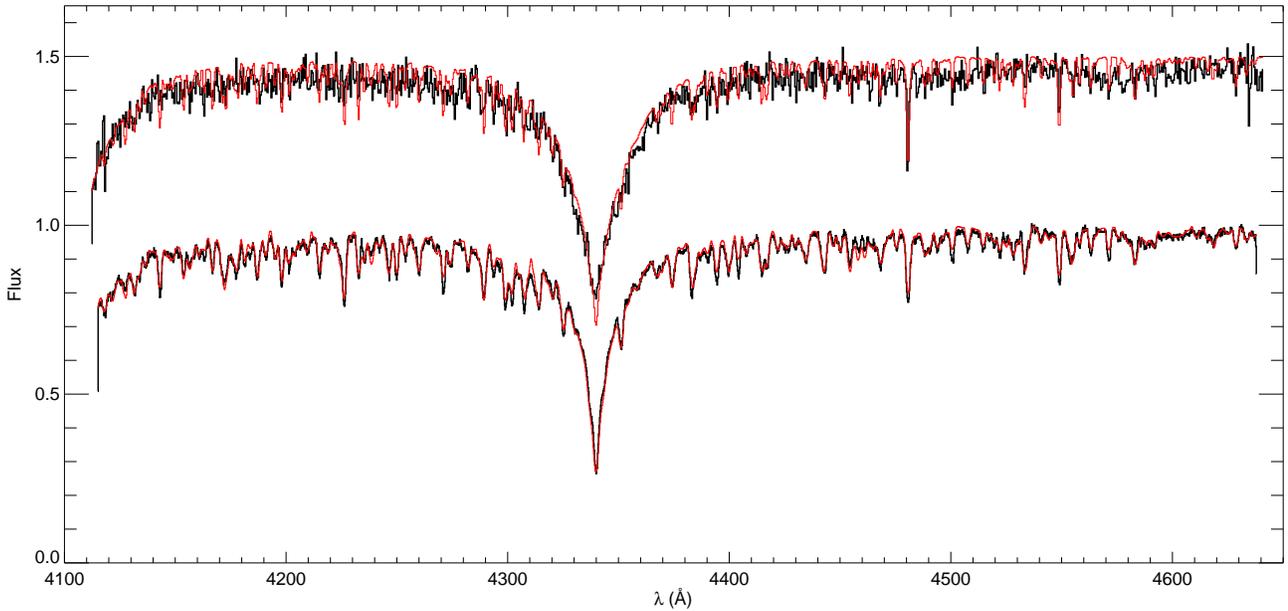}}
\caption{SME model fits (thin lines, red/grey) to the disentangled spectra
(thick lines, black) of WASP\,1628+10\,A and WASP\,1628+10\,B (offset by +0.5
units). \label{sme}} \end{figure*}

 We have used the software package {\sc sme} \citep[``Spectroscopy Made
Easy'',][]{1996A&AS..118..595V} version 412~beta to fit synthetic spectra
based on Kurucz' {\sc atlas9} model atmospheres \citep{2002A&A...392..619H} to
the disentangled spectra of WASP\,1628+10\,A and WASP\,1628+10\,B. Atomic and
molecular line data were obtained from the Vienna Atomic Line Database version
VALD3\footnote{\it vald.inasan.ru/$\sim$vald3}. The option to use the extended
treatment of van~der~Waals broadening was used \citep{1998PASA...15..336B}.
Line lists for each star were generated using the ``Extract Stellar'' option
of VALD3 to find lines with an expected depth of at least 0.01 for stars with
$\Teff =7500$\,K or $\Teff =8500$\,K,  $\log g=4.5$ and solar composition.
We fitted the entire available spectral range, which includes the red wing of
the H$\delta$ line and the entire H$\gamma$ line. The free parameters in the
fit were the radial velocity, $V_r$; the projected equatorial rotational
velocity, \Vsini; the effective temperature, $\Teff$, and the metallicity,
[Fe/H]. The surface gravity was fixed at the value $\log g=4.2$ for
WASP\,1628+10\,A and $\log g=4.5$ for WASP\,1628+10\,B. These values are
close to the values derived directly from the analysis of the lightcurves and
radial velocity data in Section~\ref{absparsect}. The micro-turbulence
parameter was fixed at the value 2\,\kms\ and the macro-turbulence parameter
was set to 0 \citep{2009A&A...503..973L}. The instrumental profile was assumed
to be a Gaussian with width corresponding to a resolving power R=10,000. The
disentangling procedure was performed such that the radial velocity defined
here is the same as the systemic velocity of the binary system. From an
initial fit to the spectra it was clear that the disentangled spectra were not
correctly normalised and so we used a low-order polynomial fit to the
residuals from this initial SME fit to re-normalize the spectra. The fits to
these re-normalized disentangled spectra are shown in Fig.~\ref{sme} and the
optimum values found by least-squares are given in Table~\ref{smeTable} with
estimated standard errors on each of the free parameters. It can be seen that
the fit to the spectrum of WASP\,1628+10\,A reproduces well the strength of
all the metal lines in the spectrum as well as the shape and depth of the
H$\gamma$ line, i.e., there is no sign that this is a chemically peculiar
star.

\begin{table} \caption{Results of SME fits to the disentangled spectra of
WASP\,1628+10\,A and WASP\,1628+10\,B. Values preceeded by ``='' were fixed
parameters in the least-squares fit. The value of $V_r$ here is consistent
with  systemic velocity, $\gamma$, as expected. \label{smeTable}}
\begin{tabular}{lrr} 
\hline
\multicolumn{1}{c}{Parameter} & \multicolumn{1}{c}{WASP\,1628+10\,A} &
\multicolumn{1}{c}{WASP\,1628+10\,B} \\ \hline \hline 
\noalign{\smallskip}
\Teff\ [K]        & 7500 $\pm$ 200    & 8650 $\pm$ 500    \\ 
\logg\ (cgs)      & =\,4.2              & =\,4.5             \\ 
$[{\rm Fe/H}]$    & $-0.3 \pm 0.3$  & $ 0.0  \pm 0.4$ \\ 
\Vsini\   [km/s] & $   105 \pm 10$    & $ 55 \pm 20$    \\ 
$V_r$ [km/s]     & $ -41   \pm 2$    & $ -40   \pm 5$    \\
\noalign{\smallskip} \hline \end{tabular}   \end{table}     

\subsubsection{Spectral energy distribution}

 We have also estimated the effective temperatures of the  stars  by comparing
the observed flux distribution to synthetic flux distributions based on the
BaSel 3.1 library of spectral energy distributions
\citep{2002A&A...381..524W}. Near-ultraviolet (NUV) and far-ultraviolet (FUV)
photometry was obtained from the GALEX GR6 catalogue\footnote{\it
galex.stsci.edu/GR6} \citep{2007ApJS..173..682M}. Optical photometry was
obtained from the NOMAD catalogue\footnote{\it www.nofs.navy.mil/data/fchpix}
\citep{2004AAS...205.4815Z}. Near-infrared photometry was obtained from the
2MASS\footnote{\it www.ipac.caltech.edu/2mass}  and DENIS\footnote{\it
cdsweb.u-strasbg.fr/denis.html} catalogues
\citep{2006AJ....131.1163S,2005yCat.2263....0T}. We used the calibration of
\citet{2014MNRAS.tmp..106C} to correct the GALEX fluxes to account for the
detector dead-time correction for bright stars. We assumed surface gravity
values of $\log {\rmn{g}}_{\rmn{A}} = 4.25$ for WASP\,1628+10\,A and $\log
\rmn{g}_{\rmn{B}} = 4.5$ for WASP\,1628+10\,B. The total line-of-sight
reddening for WASP\,1628+10 from the maps of
\citet{2011ApJ...737..103S}\footnote{\it
ned.ipac.caltech.edu/forms/calculator.html} is E(B$-$V)=0.056 and our
estimated standard error on this value is 0.034
magnitudes \citep{2014MNRAS.437.1681M}. Additional constraints included in the
fit are the observed values of the luminosity ratio, $\ell_{\rm r'}$, and  and
surface brightness ratio, $J_{\rm r'}$, from the fits to the r'-band
lightcurves given in Table~\ref{lcfitTable}. Further details of the method are
given in  \citet{2014MNRAS.437.1681M}.

\begin{figure} \mbox{\includegraphics[width=0.47\textwidth]{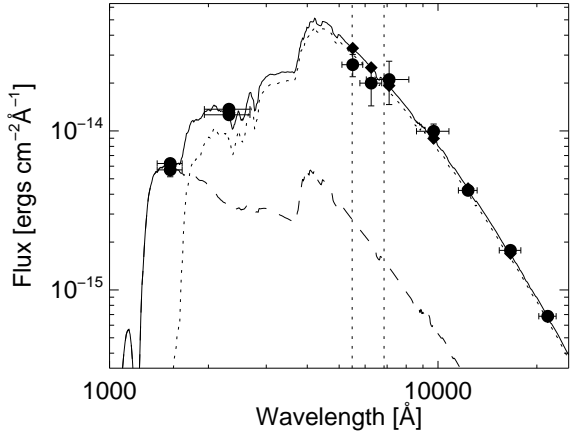}}
\caption{Model fit to the observed flux distribution of WASP\,1628+10 used to
estimate the effective temperatures of WASP\,1628+10\,B. The observed fluxes
are shown as circles with error bars. The predicted contributions of each star
to the observed fluxes are shown as dotted or dashed lines and their sum is
shown as a solid line. The models have been smoothed slightly for clarity in
these plots. Diamonds show the result of integrating the total model flux over
the band-width indicated by horizontal error bars on the observed fluxes. The
assumed band-width of the r' filter is indicated with vertical dotted lines.
The model for WASP\,1628+10\,A has $\Teff = 7400$\,K [{\rm Fe/H}]=$-0.5$ and
\logg=4.25. The model for WASP\,1628+10\,B has $\Teff =9800$\,K,[{\rm
Fe/H}]=$-0.5$ and \logg=4.5. A reddening of E(B$-$V)=0.056 has been assumed.
\label{fitflux}} \end{figure}

 Based on the results of the spectral analysis above we restrict the
comparision of the observed fluxes to models in which the metallicity of
WASP\,1628+10\,A is [{\rm Fe/H}]=$-0.5$ or [{\rm Fe/H}]=$0$.  The results are
quite sensitive to the assumed reddening to the system and so we also imposed
the value of T$_{\rm eff,A}=7500 \pm 200$\,K from the analysis of the
disentangled spectra as a constraint and only use the analysis of the spectral
energy distribution to estimate the value of T$_{\rm eff,B}$.  With these
constraints  we find T$_{\rm eff,B} = 9800\pm300$\,K. The fit to the observed
fluxes is shown in Fig.~\ref{fitflux}.

\subsection{Absolute parameters\label{absparsect}} The mass, radius, luminosity and other
parameters of WASP\,1628+10\,A and WASP\,1628+10\,B based on the results above
are given in Table~\ref{abspartable}. The adopted value of ${\rm T}_{\rm
eff,B}$ is the average of the values derived from fitting its spectrum and
from fitting the spectral energy distribution. The quoted error on this value
reflects the difference between these two estimates.

\begin{table} \caption{Physical parameters of  WASP\,1628+10\,A and
WASP\,1628+10\,B inferred from our observations. V$_{\rm synch}$ is the
equatorial rotation velocity assuming synchronous rotation.
\label{abspartable}} \begin{tabular}{lrr} \hline 
\multicolumn{1}{l}{Parameter}& 
\multicolumn{1}{c}{WASP\,1628+10\,A} &
\multicolumn{1}{c}{WASP\,1628+10\,B}\\ 
\hline \hline \noalign{\smallskip} 
Mass[\Msolar]         &$ 1.36 \pm 0.05 $&$0.135\pm 0.02   $\\ 
Radius [\Rsolar]      &$ 1.57 \pm 0.02 $&$0.348\pm 0.008  $\\ 
\Teff [K]              &$ 7500 \pm 200 $&$9200 \pm 600    $\\ 
$\log L/\Lsolar$       &$ 0.85 \pm 0.05 $&$-0.1 \pm 0.1   $\\ 
\logg (cgs)            &$ 4.18 \pm 0.01 $&$4.49 \pm 0.05   $\\ 
V$_{\rm synch}$ [\kms] &$ 110  \pm 2    $&$24   \pm 1      $\\ 
$[{\rm Fe/H}]$         & $-0.3 \pm 0.3$  & $ 0.0  \pm 0.4$ \\ 
\hline 
\end{tabular}
\end{table}

\begin{figure} \mbox{\includegraphics[width=0.47\textwidth]{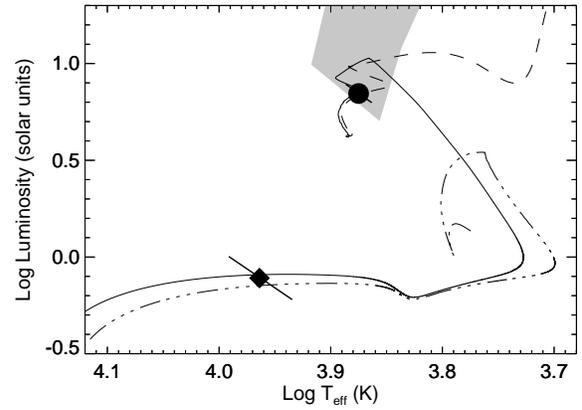}}
\caption{WASP\,1628+10\,A (circle) and WASP\,1628+10\,B (diamond) in the
Hertzsprung-Russell diagram. The dashed line is a model for  a star with a
mass of 1.36\Msolar\ and metallicity Z=0.006. The solid line is a
model for a star with an initial mass of 1.30\Msolar\ in a binary system with
a 0.8\Msolar\ companion, orbital separation of 4.8\Rsolar\  and metallicity
Z=0.004. The dot-dash line is a model for a star with an initial mass of
1.0\Msolar\ in a binary system with a 0.9\Msolar\ companion, orbital
separation of 4.8\Rsolar\  and metallicity Z=0.008. The instability strip for
$\delta$-Scuti-type pulsations with radial order $k = 4$ is indicated by
light-grey shading \citep{ 2004A&A...414L..17D}. \label{hrdfig} }
\end{figure}

\section{Discussion}

 In Fig.~\ref{hrdfig} we show WASP\,1628+10\,A  and WASP\,1628+10\,B in the
Hertzsprung-Russell diagram (HRD) compared to various models produced with the
{\sc garstec} stellar evolution code \citep{2008Ap&SS.316...99W}. We compared
WASP\,1628+10\,A to a grid of models for single stars over a finely-sampled
grid of stellar masses previously produced with {\sc garstec} for the
determination of stellar parameters from spectro-photometric and asteroseismic
data  \citep{2013MNRAS.429.3645S}. The properties of WASP\,1628+10\,A are well
matched by stellar models of the appropriate mass and with a composition
consistent with the value derived in Section~\ref{smesect}. Models with solar
composition are also a reasonable match to the observed position  of this star
on the HRD if the mass is assumed to be towards the lower limit allowed by our
observations, or if we assume an enhanced helium abundance and solar
metalicity for this star. It is likely that this star has accreted a
substantial fraction of its current mass from its companion during the
formation of the pre-He-WD. With better quality data it may be possible to
detect surface composition anomalies or other signatures of the accretion
history of this star.

 For WASP\,1628+10\,B we created a grid of models using a modified version of
{\sc garstec} that accounts for Roche lobe overflow (RLOF) in binary evolution
by forcing the radius of the star to match the radius of the Roche lobe during
the semi-detached evolutionary phase. The grid of models comprises about 3000
evolutionary tracks. The mass of the donor ranges from 1.0 to 1.5~M$_\odot$ in
steps of 0.1~M$_\odot$ for metallicity values $Z= 0.004, 0.008, 0.012, 0.016,
0.020$. The mass of the original secondary star (now WASP\,1628+10\,A) ranges
from 0.7 to 1.5~M$_\odot$ (also in steps of 0.1~M$_\odot$) and the orbital
separation at the time mass transfer begins ranges from 4 to 5~R$_\odot$ in
steps of 0.1~R$_\odot$. The orbital evolution is followed as described in
\citet{2013Natur.498..463M}, with mass loss and magnetic braking as the two
mechanisms responsible for extracting angular momentum from the system. It is
assumed that mass lost from the systems carries away all its angular momentum,
and we assume a fraction $f=0.5$ of the mass lost by the donor is accreted
onto the secondary.

 A large number of these models match the observed properties of
WASP\,1628+10\,B within the current large uncertainties on the mass and
temperature of this star. Two such models are shown in Fig.~\ref{hrdfig}, one
of which has the correct mass and composition to be a good match in its early
evolutionary phases  to the current properties of WASP\,1628+10\,A. We did not
find any models that can fit the properties of WASP\,1628+10\,B  and that
also match the orbital period of the binary -- the models predict orbital
periods less than about 0.4\,days. This may be a result of the assumptions
made in our model about the way that the star reacts to mass loss. If the star
has a larger radius than assumed during the mass loss phase then the prediced
orbital period predicted will also be increased. We have not explored this
problem further, but this will certainly be a worthwhile exercise once we have
stronger constraints on the mass, temperature and luminosity of this star.

The high-frequency signals we have detected in our Ultracam photometry are
likely to be due to pulsations in WASP\,1628+10\,B similar to those seen in
WASP\,0247$-$25\,B, i.e., a mixture of non-radial and radial overtone modes
with p-mode characteristics in the envelope and g-mode characteristics in the
interior of the star. These pulsations can enable detailed studies to be made
of the interior of a star if the exact frequencies and oscillation modes can
be identified, particularly if strong observational constraints on the mass,
temperature and luminosity of the star are available. 

 Our models for the formation of WASP\,1628+10\,B do not include microscopic
diffusion. In the models including diffusion that we computed for
WASP\,0247$-$25\,B we found that  the surface abundance of metals dropped to
almost zero by the time the effective temperature of the star had risen to
$\sim 7500$\,K due to gravitational settling. The timescale for gravitational
settling is expected to be similar for WASP\,1628+10\,B. The  presence of a
strong Mg\,II~4481\AA\ line in the spectrum of WASP\,1628+10\,B shows that
some process  counteracts or prevents gravitational settling in its
atmosphere. Radiative levitation may play a role in maintaining a metal-rich
atmosphere, but it is also possible that the contraction of this star as
it evolves to higher effective temperature produces an angular velocity
gradient that drives rotation-induced mixing. It may be possible to study this
issue  in much greater detail if it can be confirmed that the high-frequency
pulsations we have observed originate from pulsational modes in
WASP\,1628+10\,B that show rotational splitting.

\section*{Acknowledgements} Based on observations made with the William
Herschel and Isaac Newton Telescopes operated on the island of La Palma by the
Isaac Newton Group in the Spanish Observatorio del Roque de los Muchachos of
the Instituto de Astrof\'isica de Canarias.  DPM was supported by a PhD
studentship from the Science and Technology Facilites Council (STFC). This
work has made use of the VALD database, operated at Uppsala University, the
Institute of Astronomy RAS in Moscow, and the University of Vienna. AS is
partially supported by  the MICINN grant AYA2011-24704 and by the ESF
EUROCORES Program EuroGENESIS (MICINN grant EUI2009-04170). TRM and SC were
supported under a grant from the UK Science and Technology Facilities Council
(STFC), ST/L000733/1. VSD and  Ultracam are supported by the STFC. 

\bibliographystyle{mn2e}  \bibliography{wasp}

\label{lastpage}

\end{document}